\documentclass[
aps,%
12pt,%
final,%
notitlepage,%
oneside,%
onecolumn,%
nobibnotes,%
nofootinbib,%
superscriptaddress,%
noshowpacs,%
centertags]%
{revtex4}%
\usepackage{geometry}
\usepackage{graphics}

\begin{document}
\selectlanguage{english}
\title{Exclusive charmed meson pair production}

\author{\firstname{A.~V.}~\surname{Berezhnoy}}
\email{aber@ttk.ru}
\affiliation{SINP MSU, Moscow, Russia}%

\author{\firstname{A.~K.}~\surname{Likhoded}}
\email{likhoded@mx.itep.su}
\affiliation{IHEP, Protvino, Russia}%

\begin{abstract}
The experimental data of  BELLE Collaboration on 
 the exclusive charmed meson pair production in the  process 
 of monophotonic \(e^+e^-\)-annihilation 
 (\(e^+e^-\rightarrow \gamma^* \rightarrow D\bar D\))
 has been studied. It has been shown that these data is 
 described  satisfactorily in the frame work of  constituent quark model.
Our studies have demonstrated that  the central production process 
\(e^+e^-\rightarrow e^+e^-\gamma\gamma \rightarrow e^+e^-D\bar D +X\) and
 the  process of monophotonic \(e^+e^-\)-annihilation  yield 
comparable numbers of the charmed meson pairs. 
\end{abstract}

\maketitle

\section{Introduction}

The exclusive meson pair production is the unique opportunity 
to research the asymptotic behaviour of the meson form factor in the frame work
of pQCD.  The heavy meson pair production is of special interest.
In this case the effective heavy quark theory  allow us to  connect 
 the meson production processes with the meson decay processes~\cite{Grozin}. 
The asymptotic form factor behaviour is described by the  
factorized amplitude which can be represented as a wave function \( f(x,Q) \) 
multiplied by a hard interaction amplitude~\cite{Brodsky}. 
In the leading order of
 \( \alpha _{s} \) only  one gluon exchange contribute into 
 the  hard interaction amplitude and the wave function 
\( f(x,Q) \) have peak at \( x=\Lambda _{QCD}/M \), where 
\( \Lambda _{QCD} \) is a strong interaction scale, and 
 \( M \) is heavy quark mass.  The distribution of  \( f(x,Q) \) over \(x\) 
 becomes thiner with increasing heavy quark mass. 
 Due to this feature one can use the following approximation
 for the wave function:  

\begin{equation}
\label{peak}
f(x,Q)\sim \delta \left( x-\frac{m_{q}}{M}\right).
\end{equation}

In this approach the momentum fraction carried out by quark is proportional to
the quark mass:\[
x_{i}\simeq \frac{m_{i}}{M}.\]

The models of such kind~\cite{Brodsky} are taking into account the hard "tail"
of the wave function only and neglect the soft component of the wave function. 
Nevertheless these models correctly reproduce the cross section 
behaviour near the threshould
as well as asymptotic behaviour at  \( Q^{2}\rightarrow \infty  \). 
Moreover, difference between the peak approximation calculation and 
the calculation beyond the peak approximation~(\ref{peak}) 
is not essential~\cite{Ji}.
In this work we endeavour to study the recent experimental results of BELLE
collaboration~\cite{Uglov} for  \( D^{*+}D^{*-} \)-pair production in 
 \( e^{+}e^{-} \)-annihilation at  \( \sqrt{s}=10.6 \)~GeV. Also we
 predict the cross section value of the charmed meson pair 
 production in  \( \gamma \gamma  \)-interaction at BELLE.

\section{\protect\( e^{+}e^{-}\protect \)-annihilation}

The following cross section values  of the charmed meson pair production
in \( e^{+}e^{-} \)-annihilation have been  
measured by BELLE~Collaboration\cite{Uglov}:\[
\sigma (e^{+}e^{-}\rightarrow D^{*+}D^{*-})=0.65\pm 0.04\pm 0.07\;
\textrm{pb},\]
 \[
\sigma (e^{+}e^{-}\rightarrow D^{+}D^{*-})=0.71\pm 0.05\pm 0.09\;
\textrm{pb}.\]
The differential cross section distributions over azimuthal angle have been 
measured too. 

\begin{figure}
{\centering \resizebox*{0.8\textwidth}{!}{\includegraphics{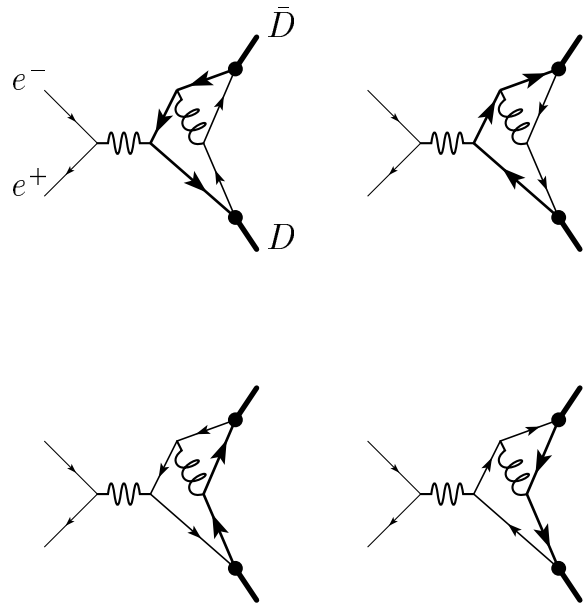}} \par}
\setcaptionmargin{5mm}
\onelinecaptionsfalse
\captionstyle{normal}
\caption{Feynman diagrams for the process of charmed meson pair production
in  \protect\( e^{-}e^{+}\protect \)-annihilation.\label{ee_diagr}}
\end{figure}

The interaction energy at BELLE (10.6~GeV) is large enough  
to determinate the leading asymptotic contribution into the meson form factor:
\[
F(Q^{2})\simeq \frac{\alpha _{s}f_{M}^{2}}{Q^{2}},\]
where \( f_{M} \) is the constant of leptonic decay, 
and \( Q^{2}=s=(k_{e^{+}}+k_{e^{-}})^{2} \)
is the  \( e^{+}e^{-} \)-interaction energy squared. Nevertheless,
 it is worth to mention that the virtuality of gluon splited into
 quark-antiquark pair \( q\bar{q} \) is slightly smaller than this energy
\[
q^{2}\sim x_{q}^{2}Q^{2},\]
 where \( x_{q} \) is the momentum fraction of the meson carried out by the
 light quark (See Fig.~\ref{ee_diagr}).
If the pre-asymptotic terms 
would be large, then the model prediction would not reliable. But as it was 
shown in paper~\cite{Ji}, the pre-asymptotic terms are not essential.

Let us return to the discussion of the constituent quark model. As it mentioned
above, the cross section behaviour near the threshould is described correctly
by this model: \( \sigma \sim \left( \frac{s}{4}-M^{2}\right) ^{(2L+1)/2} \),
where \( L \) is orbital momentum of the meson pair. (For production in
\(e^+e^-\)-annihilation \(L=1\).)
At \( M\rightarrow \infty  \) the constituent quark model, as well as 
the effective theory of heavy quarks give the same predictions for  
the cross section ratio:
\[\sigma _{PP}:\sigma _{PV}:\sigma _{VV}=1:4:7,\]
where \( P \) and \( V \) are pseudoscalar state and vector state,
correspondingly~\cite{Kiselev}.
At large energy  the cross sections \(\sigma _{PP}\),
\(\sigma _{PV}\) and  \(\sigma _{VV}\) behave as follows:  
\[\sigma \sim \frac{\alpha ^{2}\alpha ^{2}_{s}}{s^{3}}.\]

An angular distribution of the cross sections have the following simple form: 
\[
\frac{d\sigma (D\bar{D})}{d\cos \Theta }\sim \sin ^{2}\Theta ,\]
\[
\frac{d\sigma (D\bar{D}^{*})}{d\cos \Theta }\sim 1+\cos ^{2}\Theta, \]
where \( \Theta  \) is the angle between the final meson direction and the 
initial beam direction in the center-of-mass system.  

For the case of \( D\bar{D}^{*} \) pair production only transversal 
component of \( D^{*} \) contributes into cross section, because the 
production of longitudinal  \( D^{*} \) component 
do not allowed by lows of parity and  angular momentum conservation.
Indeed, in the center-of-mass system
the matrix element of \( D\bar{D}^{*} \) production 
in \( 1^{-} \) state looks like follows:
\[
[\mathbf{\epsilon_{D^*}}(\mathbf{p})\times \mathbf{p}]\cdot 
\mathbf{\epsilon_q}\varphi_{D},\]
where \( \mathbf{\epsilon_{D^*}} \) and \( \mathbf{\epsilon_q} \)
are  polarizations of \( D^{*} \) and virtual photon, correspondingly. 
One can see from this formula, that the longitudinal component contribution
 \( D^{*} \) \[
\mathbf{\epsilon_{\parallel}}=\mathbf{n}(\mathbf{n}\epsilon_{D^{*}})
\frac{E}{m},\]
equals to zero ( \( \mathbf{n}=\mathbf{p}/|\mathbf{p}| \), 
 \( E \) and \( M \) are energy and mass of \( D^{*} \) meson).
 
The longitudinally polarized \( D^{*} \) meson could be produced 
in the process of diphotonic \( e^{+}e^{-} \)-annihilation, 
but as is was shown in ~\cite{Liu}, the charmed meson yield 
 is small for this process. 
 
The pair production of vector mesons  \( D^*\) and \(\bar{D}^{*} \)
in the center-of-mass system can  be describe by two independent 
structures:
\[
M_{1}\sim (\mathbf{\epsilon_1}\mathbf{p})(\mathbf{\epsilon_2}
\mathbf{\epsilon_q})
+(\mathbf{\epsilon_2}\mathbf{p})(\mathbf{\epsilon_1}\mathbf{\epsilon_q}),\]

\[
M_{2}\sim 
(\mathbf{\epsilon_1}\mathbf{\epsilon_2})(\mathbf{p}\mathbf{\epsilon_q}).\]

Both amplitudes correspond to \(1^-\) state of the dimeson system.
First amplitude contribute to \( D_{L}^{*}D^{*}_{T} \)-pair production.
Second one contribute to all possible variants of particle polarizations.
The ratio between these amplitudes depends on the quark masses.

The satisfactory description of 
the experimental data of BELLE Collaboration 
has been achieved for the following values of the model parameters: 
\[
\alpha _{s}=0.3,\]
\[
f_{D}=200\;\textrm{MeV},\]
\[
m_{q}=0.17\;\textrm{GeV},\]
\[
m_{c}=1.5\;\textrm{GeV}.\]
We have predicted the following cross section values:
\[
\sigma (D^{*+}D^{*-}):\sigma (D^{+}D^{*-}):\sigma (D^{+}D^{-})=
0.73\;\textrm{pb}:0.58\;\textrm{pb}:0.02\;\textrm{pb}\]

One can see that the production of two pseudoscalar mesons is suppressed by the
order of magnitude.   

\begin{figure}
{\( d\sigma /d\cos \Theta  \), pb \hfill}

{\centering \resizebox*{\textwidth}{!}{\includegraphics{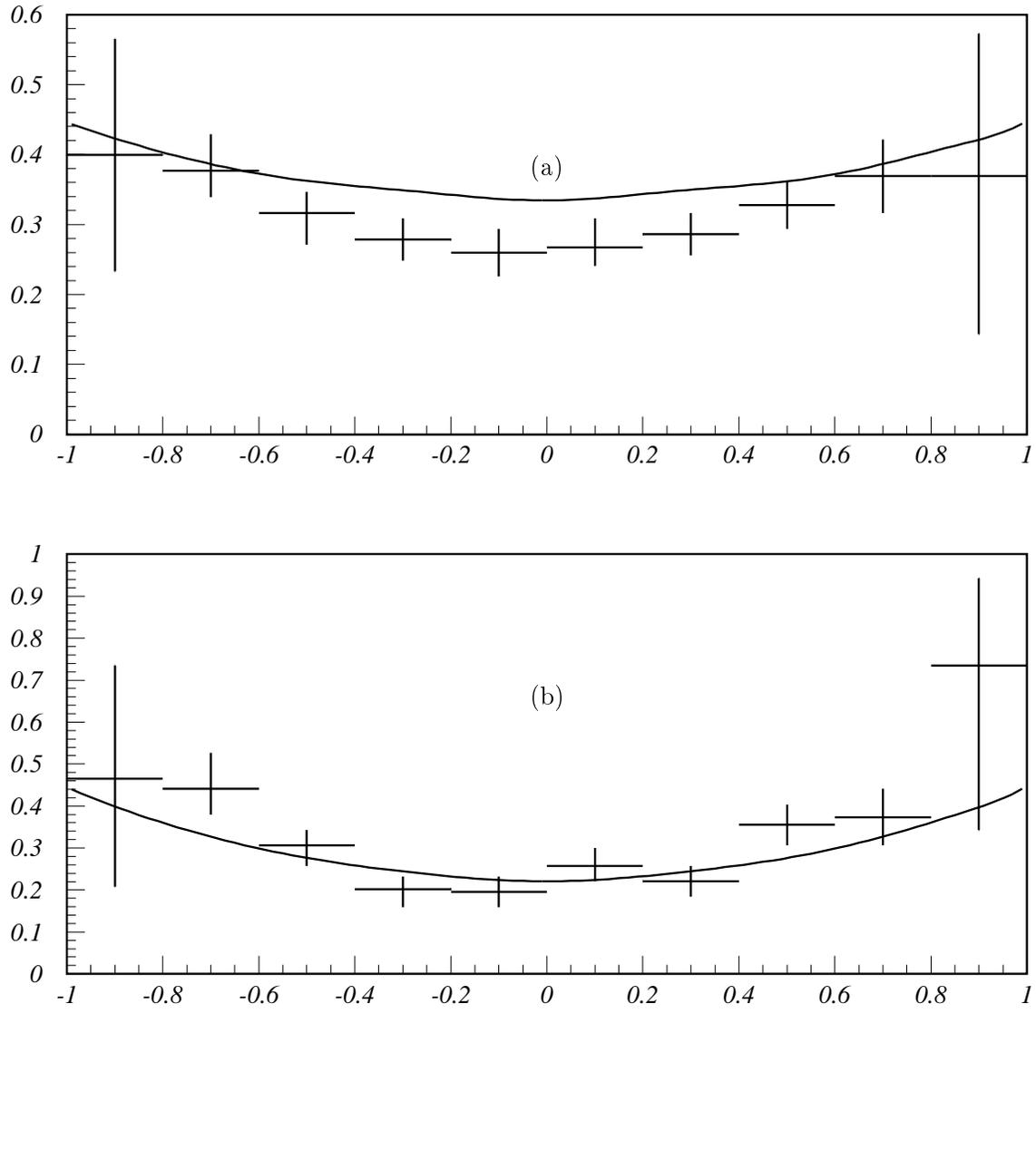}} \par}

{\hfill \( \cos \Theta  \)}

\vspace*{-15.cm}\centering{(a)}

\vspace*{7.0cm}\centering{(b)}

\vspace*{6.3cm}

\setcaptionmargin{5mm}
\onelinecaptionsfalse
\captionstyle{normal}
\caption{The cross section distribution over  
\protect\( \cos \Theta \protect \)~(pb) 
in comparison with the experimental data for the process
\protect\( e^{+}e^{-}\rightarrow D^{*}\bar{D}^{*}\protect \) (a) and
for the process  
 \protect\( e^{+}e^{-}\rightarrow D^{*}\bar{D}\protect \) (b).
\protect\( m_{c}=1.5\protect \) GeV, \protect\( m_{q}=0.17\protect \)
GeV, \protect\( f_{D}=200\protect \) MeV, \protect\( \alpha _{s}=0.3\protect
\).\hfill \label{ee_theta}}
\end{figure}

The experimental angular cross section distribution \( d\sigma /d\cos \Theta  \)
for  the processes of \( D^{*}\bar{D}^{*} \)- and \( D\bar{D}^{*} \)-pair
production in comparison with the model predictions has been shown
in Fig.~\ref{ee_theta}. The best description of the
experimental data  is 
achieved for  \( m_{q}=0.17 \)~GeV and \( m_{c}=1.5 \)~GeV.

It is worth to mention that the model under discussion is rather rough
approach, which takes into account only hard ``tail'' of 
the meson wave function. 
The contribution of the soft part of wave function has been neglected. 
That is why the cross section dependence on energy have not minimum for
the case of production of the two pseudocalar mesons, which has been predicted
in the frame work of the  effective theory of heavy quarks~\cite{Grozin}.
However cross section values predicted in~\cite{Grozin} are large than
experimental data by the order of magnitude. 
It is worth to mention ones more,  that the   results  of  calculations above 
the peak approximation~(\ref{peak}) do not differ essentially from
 the results obtained in the frame work of peak approximation~\cite{Ji}.

\section{Photonic production of  \protect\( D^{(*)}\bar{D}^{(*)}\protect
\)-pair }

\begin{figure}
\centering{ \resizebox*{0.9\textwidth}{!}{\includegraphics{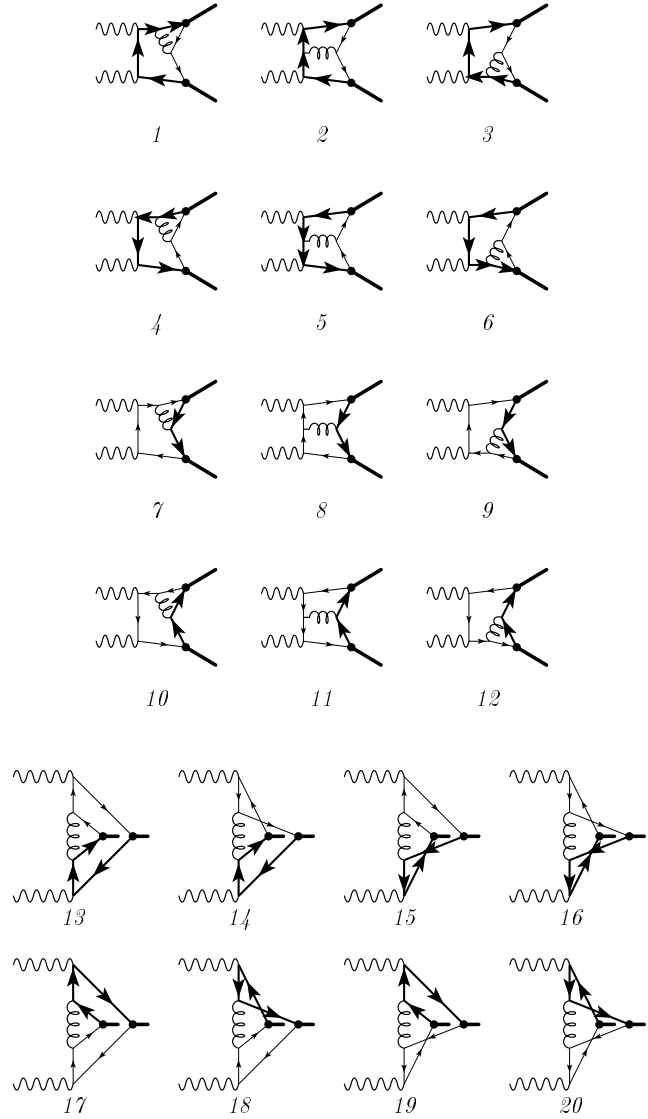}} }

\setcaptionmargin{5mm}
\onelinecaptionsfalse
\captionstyle{normal}
\caption{Feynman diagrams for the charmed meson pair production in the 
photon-photon interaction. \hfill \label{phph_diagr}}
\end{figure}

The exclusive production of  \( D^{(*)}\bar{D}^{(*)} \)-pair in the
photon-photon interaction has been studied in our paper~\cite{BKL}.
Our results have been compared with the predictions of the
effective theory of heavy quarks. In the constituent quark approach
the photonic production of the charmed meson pair is described by
twenty  Feynman tree level diagrams (see Fig.~\ref{phph_diagr}). 
The analogous diagram set has been used to calculate the inclusive 
 \( B_{c} \)-meson production cross section~\cite{BSL}. 
This diagram set can be subdivided into three 
gauge invariant parts.
The first group of diagrams  (1-6 of Fig.~\ref{phph_diagr}) corresponds to the 
case when  heavy quark  radiates gluon which splits into light quark pair.
The second group ( diagrams 7-12) can be received from the first one by  
permutation  \( Q\leftrightarrow q \). 
The third diagram group (diagrams 13-20 of Fig.~~\ref{phph_diagr}) 
corresponds to the independent production of the 
of the quark pairs ( \( \gamma \rightarrow QQ \), 
\( \gamma \rightarrow q\bar{q} \)) followed by there fusion into mesons.

The third diagram group dominates at relatively  small transverse momenta.
Due to the contribution of such diagrams the factorization theorem
can not be applied to describe the inclusive production of \(B_c\)-meson in 
the wide kinematic region~\cite{BSL}.
In particular, the strong cross section dependence on 
light quark charge \( e_{q} \) exists near the threshould as well as
at large interaction energy. Also, we have observed 
  such \( e_{q} \) dependence for 
  the process of photonic \( D^{(*)}\bar{D}^{(*)} \) pair
production. That process feature  has underlined also 
in the paper~\cite{Baek} for the region of large energies.  

In this article we continue our study  of the constituent 
quark model prediction for different kinematic regions.
Our analysis is based on diagrams of Fig.~\ref{phph_diagr}. 

\begin{figure}
{\( \sigma  \),~pb \hfill}

{\centering \resizebox*{\textwidth}{!}{\includegraphics{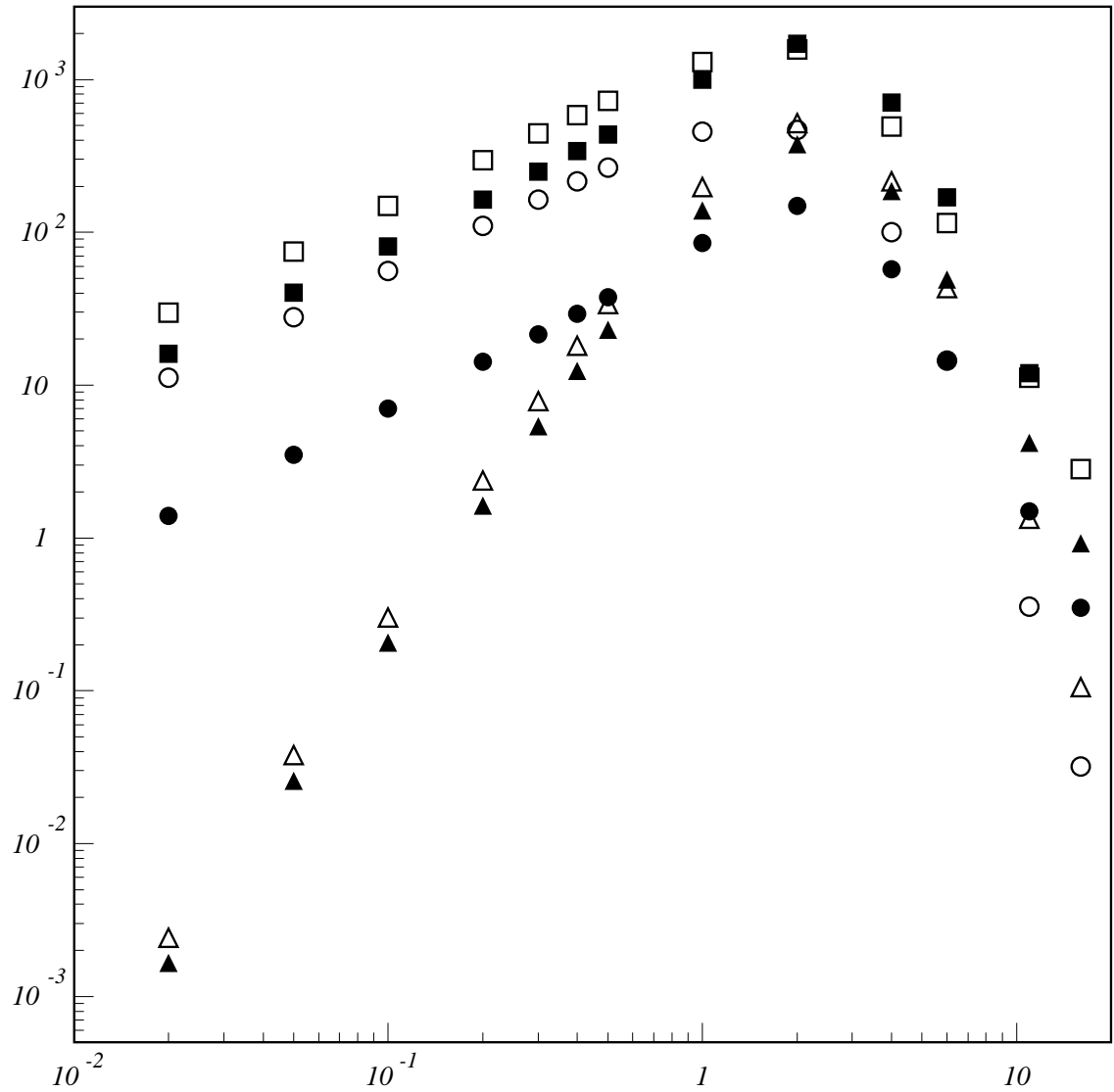}} \par}

{\hfill \( k \), GeV}

\setcaptionmargin{5mm}
\onelinecaptionsfalse
\captionstyle{normal}
\caption{The cross sections 
\protect\( \sigma _{PP}\protect\;(\circ,\:\bullet) \),
\protect\( \sigma _{PV}\protect\;(\triangle,\:\blacktriangle) \), 
\protect\( \sigma _{VV}\protect\;(\square,\:\blacksquare)\)
 as a function of \protect\( k=\sqrt{s-s_{th}}\protect \) for  
neutral \protect\( D\protect \)-mesons (closed symbols) and for 
charged \protect\( D\protect \)-mesons (opened symbols). 
\hfill \label{th_DD}}
\end{figure}

Let us consider the \( D^{(*)}\bar{D}^{(*)} \) pair production near
the  threshould.
As it has been mentioned above, \( D^{*}\bar{D}^{*} \) pairs
and  \( D\bar{D} \) ones are  produced  near the threshould 
in  \( S \)-wave state.  (The cross section depends on momentum \(k\)
between \( D \)-mesons in the center-of-mass system
as \( \sim k^{2L+1} \)).  This state corresponds to the 
total momentum and parity of the diphoton system  \( 0^{+} \). 
On the contrary, 
\( D^{*}\bar{D} \)-pair is produced in  \( P \)-wave, because  
\( S \)-wave state production is not allowed by the Landau-Yang theorem.

 \( S \)-wave state of the meson pair is produced 
 from the  \( \gamma \gamma  \) system \( 1^{+} \).
\( P \)-wave state of the meson pair \( D^{*}\bar{D} \) corresponds to
the  \( \gamma \gamma  \) system \( 0^{-} \), thus the cross 
section of the  \( D^{*}\bar{D} \)  production increases near threshold as 
\( k^{3} \). As one can see in Fig.~\ref{th_DD}, 
the cross section values of \( D^{(*)}\bar{D}^{(*)} \)-pair production
depend on light quark charge, whereas quantum numbers of the meson pair,
as well as energy behaviour of the cross sections
do not depend on  this charge value.
Also, one can see, that the cross section distribution over angle 
for neutral meson pair and charged meson pair  
differ from each other considerably.  

\begin{figure}
{\( d\sigma /d\cos \Theta  \), pb \hfill}

{\centering \resizebox*{0.8\textwidth}{!}{\includegraphics{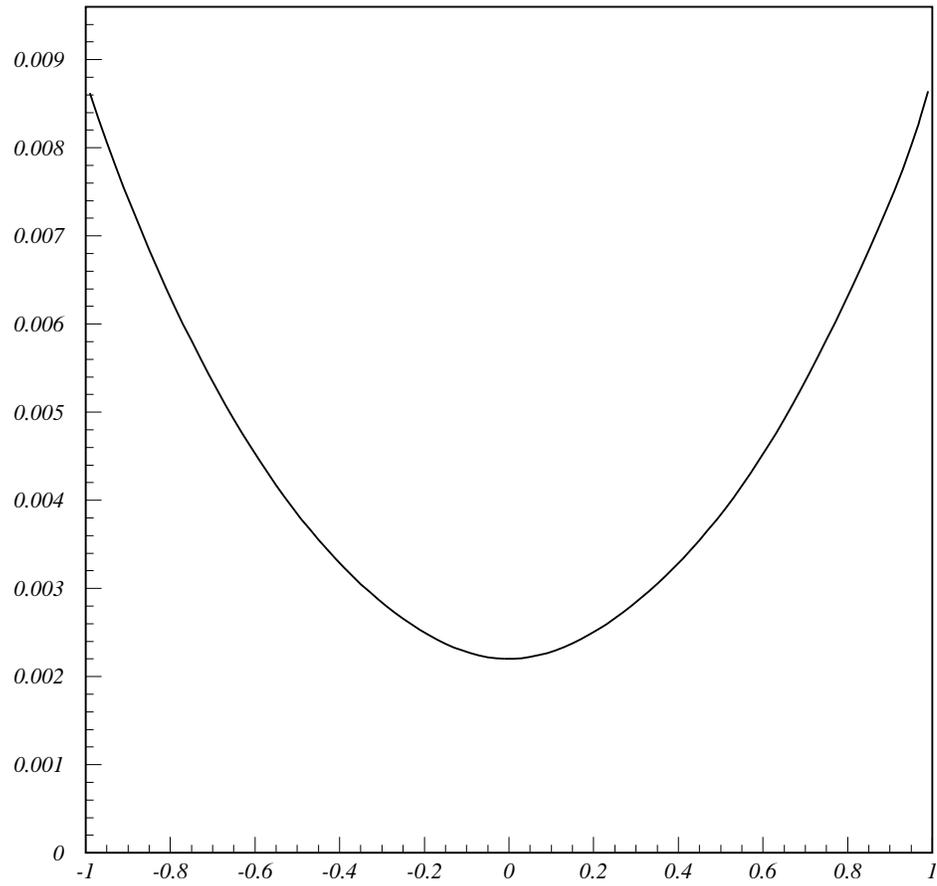}} \par}

{\hfill \( \cos \Theta  \)}

\setcaptionmargin{5mm}
\onelinecaptionsfalse
\captionstyle{normal}
\caption{The cross section distribution over 
\protect\( \cos\Theta\protect \) for the processes of 
\protect\( D^{+}D^{*-}\protect \)
and \protect\( D^{-}D^{*+}\protect \) pair production in the 
× \protect\( e^{+}e^{-}\protect \)-annihilation
at \protect\( k=\sqrt{s-s_{th}}=0.1\protect \) GeV. (Other meson pairs
are produced at small \protect\( k\protect \) isotropically.) 
\hfill\label{dd_pv}}
\end{figure}

Our model predicts practically isotropic production of 
the  \( D^{0}\bar{D}^{0} \)-,
\( D^{*0}\bar{D}^{0} \)-, \( D^{0}\bar{D}^{*0} \)-, \( D^{+}D^{-} \)-
and \( D^{*+}D^{*-} \)-pairs near the threshould: \[
\frac{d\sigma }{d\cos \Theta }\approx const,\]
whereas  \( D^{+}D^{*-} \)-pair produce peripherally even near the threshould
 (see Fig.~\ref{dd_pv}). However, as it can be clearly seen in Fig.~\ref{th_DD},
 the production of such pairs is suppressed near the threshould.
It is worth to mention, that the total cross section value of the 
neutral meson production is three times large than the   cross section value
of the charged meson production.  

At large energies the cross section distributions over \( \cos \Theta  \) 
have peripheral form with striking maxima at +1 and -1. 
This fact has been also emphasized in the  paper~\cite{Baek}, where 
the analytic form of amplitude  of the pseudocalar meson pair production 
has been performed  at hight energy limit: 
  \[
A^{\textrm{PP}}\sim \left[ (e_{Q}-e_{q})^{2}\frac{1+\beta
^{2}\cos\Theta^{2}}{1-\beta ^{2}\cos\Theta^{2}}+2e_Q^2\right] ,\]
 where \[
\displaystyle \beta =\sqrt{1-\frac{4m^{2}}{s}}.\]
 At large value of  \( s \) one can see the difference between the 
 asymptotic behaviour
of the cross sections for the charged and neutral meson production. 
The production cross section depends on 
\(s\) as \( 1/s^{2} \) for all charged meson pair.  On the contrary,
the production cross sections for neutral meson pairs have different
asymptotes:   
 \[
\sigma _{D^{0}\bar{D}^{0}}:\sigma _{D^{*0}\bar{D}^{0}}:
\sigma _{D^{*0}\bar{D}^{*0}}\sim \frac{1}{s^{3}}:
\frac{1}{s^{4}}:\frac{1}{s^{2}}.\]

However, it is worth to mention that 
the  cross section behaviour   becomes asymptotic after  \( \sqrt{s}>20 \) GeV, 
therefore the study of it have  the theoretical importance only.
In addition,  the next order logarithmic corrections can play essential role
at such energies.

\section{\protect\( D^{(*)}\bar{D}^{(*)}\protect \)-pair production 
in \protect\(\gamma\gamma \protect \)-interaction at BELLE}

It has been shown in the paper~\cite{Liu} that the contribution 
of diphotonic annihilation  
\begin{equation}
\label{gammavirt}
e^{+}e^{-}\rightarrow \gamma ^{*}\gamma ^{*}\rightarrow D\bar{D}
\end{equation}
into the total cross section is small. 

Nevertheless, another opportunity to produce  charmed meson pairs in the 
photon-photon interaction exists.  Charmed mesons could be produced 
in the interaction between the effective (equivalent) photons 
radiated by the initial fermions:

\begin{equation}
\label{gammareal}
e^{+}e^{-}\rightarrow e^{+}\gamma e^{-}\gamma \rightarrow D\bar{D}+e^{+}+e^{-}.
\end{equation}

The cross section of this process is described in the frame work of the
``partonic'' model: \[
\sigma =\int \int \sigma _{\gamma \gamma }(\hat{s})f^{\gamma }(x_{1})f^{\gamma }
(x_{2})dx_{1}dx_{2},\]
where  \( x_{1} \) and \( x_{2} \) are momentum fractions of the initial
electron and positron, correspondingly, carried out by interacting photons.  
The photonic density \( f^{\gamma }(x) \) obeys Weiz\"ecker-Williams formula: 

\[
f^{\gamma }(x)=\frac{\alpha _{\textrm{em}}}{2\pi x}\left( \left( 1-(1-x)^{2}\right) 
\ln \frac{Q^{2}_{\textrm{max}}}{Q^{2}_{\textrm{min}}}-m^{2}_{e}x^{2}
\left( \frac{1}{Q_{\textrm{min}}}
-\frac{1}{Q_{\textrm{max}}}\right) \right) ,\]
where \( m_{e} \) is the mass of electron, 
\( Q^{2}_{\textrm{min}}=m^{2}_{e}x^{2}/(1-x) \),
and \( Q_{\textrm{max}}^{2}\simeq 1 \) GeV\( ^{2} \). 

In contrast to the diphotonic annihilation~(\ref{gammavirt}) the amplitude 
of process~(\ref{gammareal}) is not suppressed by \( s \)-canal propagators.
Thus it would be interesting to estimate the cross section value for the process
of the charmed meson production in the interaction between the effective photons. 

For the parameter values used to estimate the cross section in 
\( e^{+}e^{-} \)-annihilation we have received the following unexpected
results for the  process~(\ref{gammareal}):
\[
\sigma (D^{*+}D^{*-}):\sigma (D^{+}D^{*-}):
\sigma (D^{+}D^{-})={1.52\;\textrm{pb}:0.33\;\textrm{pb}:0.13\;
\textrm{pb}.}\]
\[
\sigma (D^{*0}\bar{D}^{*0}):\sigma (D^{0}\bar{D}^{*0}):
\sigma (D^{0}\bar{D}^{0})=
1.39\;\textrm{pb}:0.43\;\textrm{pb}:0.40\;\textrm{pb}.\]

\begin{figure}
{\( d\sigma /d\cos \Theta  \), pb \hfill}

{\centering \resizebox*{\textwidth}{!}
{\includegraphics{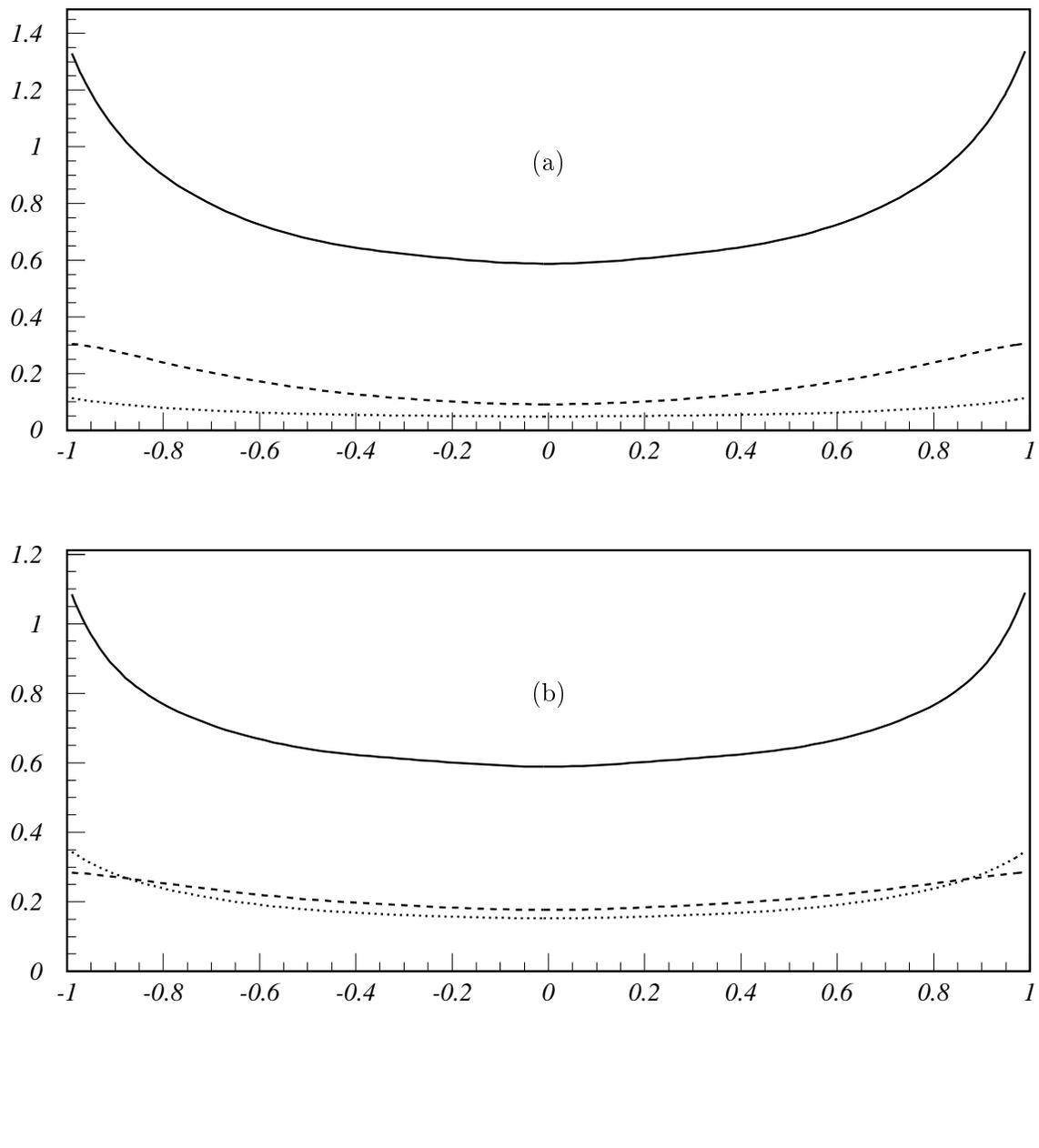}} \par}

{\hfill \( \cos \Theta  \)}

\vspace*{-15.cm}\centering{(a)}

\vspace*{7.0cm}\centering{(b)}

\vspace*{6.3cm}

\setcaptionmargin{5mm}
\onelinecaptionsfalse
\captionstyle{normal}
\caption{The differential cross section 
\protect\( d\sigma_{VV}/d\cos \Theta \protect \) (solid curve),
\protect\( d\sigma_{PV}/d\cos \Theta \protect \) (dashed curve) and
\protect\( d\sigma_{PP}/d\cos \Theta \protect \) (dotted curve)
for the processes of charged meson production (a) and 
neutral meson production in the effective photons interaction. 
 \hfill\label{eephoton_theta} }
\end{figure}

\begin{figure}
{\( d\sigma /dz, \) pb \hfill} 

{\centering \resizebox*{\textwidth}{!}{\includegraphics{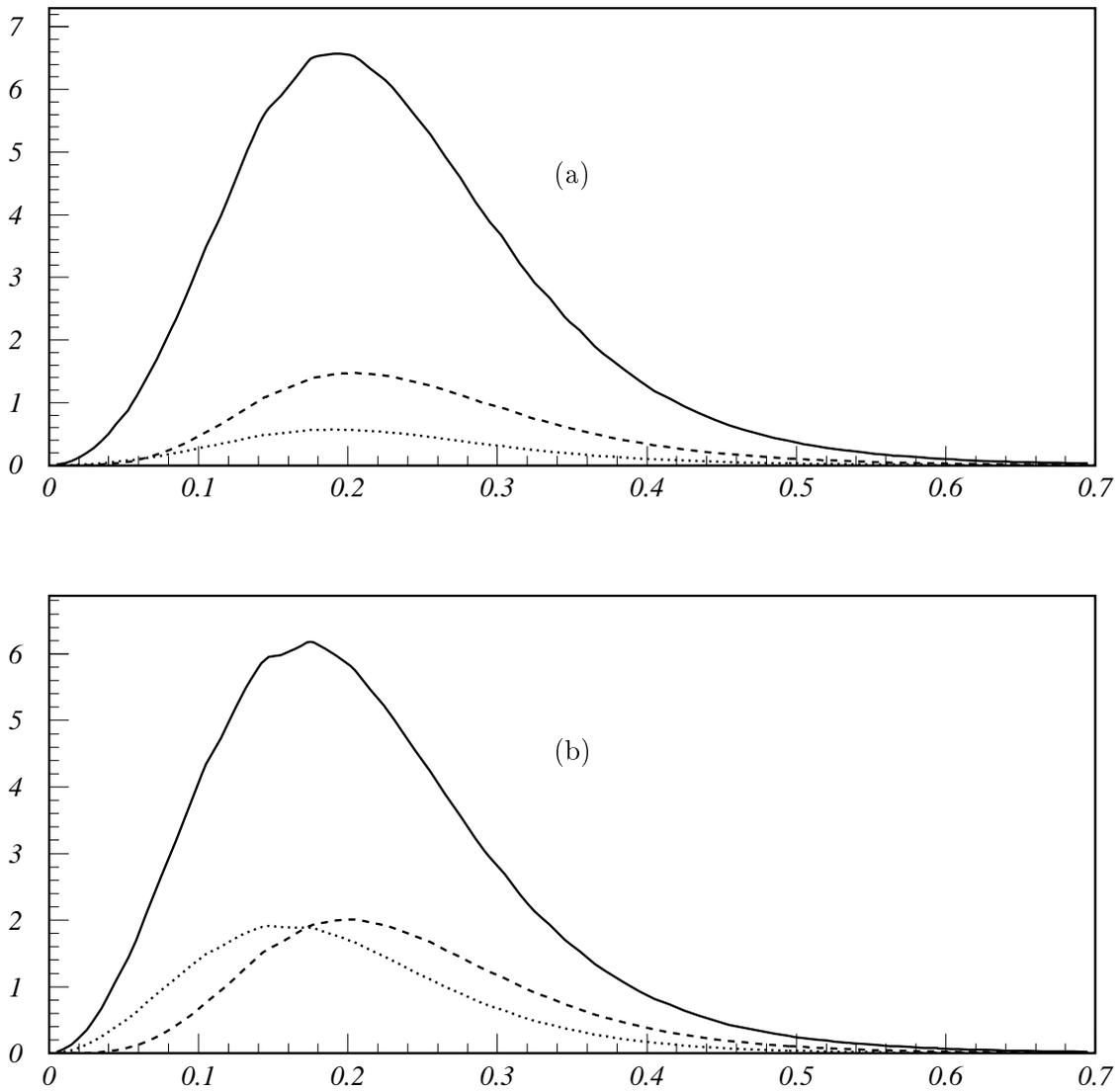}} \par}

{\hfill \( z \)}

\vspace*{-15.cm}\centering{(a)}

\vspace*{7.0cm}\centering{(b)}

\vspace*{6.3cm}

\setcaptionmargin{5mm}
\onelinecaptionsfalse
\captionstyle{normal}
\caption{The cross section distributions over 
\protect\( z=\frac{2|\mathbf{p_D}|}{\sqrt{s}} \protect \)
for the processes of charged meson production (a) and 
neutral meson production (b) in the effective photon interaction. 
The  designations are the same as in Fig.~\ref{eephoton_theta}. 
\hfill \label{effective_z}}
\end{figure}

It is clear that these cross section values are comparable with ones for
the process of  \( e^{+}e^{-} \)-annihilation. One can see in 
Fig.~\ref{eephoton_theta},
that the distribution shapes do not differ essentially from
ones for the case of \( e^{+}e^{-} \)-annihilation. 
It is obvious, that the meson energy is less than the energy of meson 
produced in \( e^{+}e^{-} \)-annihilation.
The distribution over 
\( z=\frac{2|\mathbf{p_D}|}{\sqrt{s}} \) is shown 
in Fig.~\ref{effective_z}, where  \(\mathbf{p_D} \) is 
the \( D \) meson momentum in the center-of-mass system of 
\( e^{+}e^{-} \)-pair.
One can see from that distribution that the averaged value of \( z \) 
is about 0.3 for the such
process. Therefore the averaged meson momentum is about 1~GeV.

It is worth to notice that the model under discussion
predicts the dominance of the vector-vector pair production.  

\section*{Conclusion}
We have shown that the experimental data on 
charmed meson pair production at BELLE (\(\sqrt{s}=10.6\)~GeV)
is described satisfactorily in the frame work of  the constituent quark model.
In that approach the hard part of the amplitude can be calculated with the help
of pQCD motivated diagrams.

 The production mechanism in \(e^+e^-\)-annihilation is rather simple.
 The virtual photon splits into  \(c\bar c\)-pair followed by  
hardonization process. Thus the cross section value do not depend on light
quark charge.

In contrast to \(e^+e^-\)-annihilation in \(\gamma\gamma\)-interaction
the cross section values depends strongly on  the light quark charge at low
energy, as well as at high energy. 
The essence of matter is that in the process   
\(\gamma\gamma\rightarrow D\bar D\) the light quark interaction with the
photonic field can not be neglected.

It has been shown that the cross section value of the 
central \(D\) meson production 
in the process
\[e^+e^-\rightarrow e^+e^-\gamma\gamma \rightarrow e^+e^-D\bar D +X\]
and the cross section value of \(D\) meson production 
in \(e^+e^-\)-annihilation are comparable.

\begin{acknowledgments}
This research is partially supported  by 
Grants of RF Education Ministry   E02-3.1-96,
CRDF M0-011-0 and RFBR 04-02-17530, and is realized 
in the frame work of scientific school  SC~1303.2003.2. 
\end{acknowledgments}

\end{document}